\documentclass[a4paper,11pt]{article}
\usepackage{pos}
\usepackage{tikz}
\usepackage{booktabs}
\usepackage{bm}
\usepackage{pos}
\usepackage{tikz}
\usepackage{booktabs}
\usepackage{bm}
\usepackage{comment}

\title{Spin-spin entanglement at high energy}
\ShortTitle{Spin-spin entanglement at high energy}

\author*[a]{Michael Fucilla}

\affiliation[a]{National Centre for Nuclear Research, Pasteura 7, 02-093 Warsaw, Poland}

\author[b,c]{Yoshitaka Hatta  }

\affiliation[b]{Physics Department, Brookhaven National Laboratory, Upton, NY 11973, USA}
\affiliation[c]{RIKEN BNL Research Center, Brookhaven National Laboratory, Upton, NY 11973, USA}

\author[d,e]{Bo-Wen Xiao}
\affiliation[d]{School of Science and Engineering, The Chinese University of Hong Kong (Shenzhen),
Longgang, Shenzhen, Guangdong, 518172, P.R. China} 
\affiliation[e]{Southern Center for Nuclear-Science Theory (SCNT), Institute of Modern Physics,
Chinese Academy of Sciences, Huizhou, Guangdong 516000, China}

\emailAdd{michael.fucilla@ncbj.gov.pl}

\abstract{Spin correlations offer a quantum-information perspective on the partonic final states produced in high-energy scattering. We discuss the spin-density matrix of a heavy quark-antiquark pair in two complementary small-$x$ processes. In coherent diffractive production, color-singlet exchange enforces an unusually strong relation between entanglement and Bell nonlocality: a longitudinal photon creates a maximally entangled pair, whereas for a transverse photon the pair is generically both entangled and Bell nonlocal, with a model-independent point of maximal entanglement. In inclusive back-to-back production, the density matrix factorizes into a hard spin tensor and the unpolarized and linearly polarized Weizs\"acker--Williams gluon distributions. The latter generates an azimuthal modulation and can increase the concurrence when the dijet relative momentum and imbalance are approximately orthogonal. Strikingly, in the saturation model considered, nonlinear effects wash out this modulation for $q_\perp\lesssim 3Q_s$, whereas the dilute BFKL limit, in which $G_2/G_0\to1$, yields a maximal modulation independent of the imbalance magnitude. These results connect quantum-information
observables with the Pomeron and saturation physics.}

\FullConference{The 33rd International Workshop on Deep Inelastic Scattering and Related Subjects (DIS2026)\\
4--8 May 2026\\
Bologna, Italy}

\newcommand{\qqbar}{q\bar q}
\newcommand{\kk}{\bm{k}}
\newcommand{\PP}{\bm{P}}
\newcommand{\qq}{\bm{q}}

\begin{document}
\maketitle

\section{Quantum correlations in high-energy QCD}

Recent evidence for spin entanglement, inferred from measurements of spin
correlations in top-quark pair production by ATLAS and
CMS~\cite{ATLAS:2023fsd,CMS:2024pts}, has turned a largely theoretical subject
into a concrete experimental program at colliders. The broader program, reviewed for
example in Ref.~\cite{Barr:2024djo}, asks which quantum states are produced by
a given hard process and which features of the underlying interaction survive
in measurable angular correlations.  Deep-inelastic scattering provides a
particularly attractive setting. In particular, the forthcoming Electron-Ion Collider (EIC) will combine high luminosity with access to the high-energy, or small-$x$, regime of QCD~\cite{AbdulKhalek:2021gbh}, while ultraperipheral collisions (UPCs) provide a complementary opportunity to probe the same dynamics  at the LHC.

We consider here quantum correlations of two processes that can be investigated at the EIC or LHC in UPCs: i.) Coherent diffraction,
$\gamma^*+p(A)\to \qqbar+p(A)$, which isolates color-singlet exchange and therefore
probes the quantum-information imprint of the Pomeron at high-energy. ii.) Inclusive dijet production,
$\gamma^*+A\to \qqbar+X$, which has a larger rate and, in the back-to-back limit, provides direct sensitivity to transverse-momentum-dependent gluon
distributions.  The results summarized here are based on
Refs.~\cite{Qi:2025onf,Fucilla:2025kit,Fucilla:2026mkg}.

For an unpolarized initial state, at the leading-order accuracy, parity invariance implies vanishing single-spin polarizations. The normalized two-qubit density matrix may therefore be written as
\begin{equation}
 \rho=\frac14\left[\mathbb I\otimes\mathbb I+
 C_{ab}\,\sigma^a\otimes\sigma^b\right],
 \qquad a,b\in\{r,n,k\},                                      \label{eq:rho}
\end{equation}
where $\hat{k}$ is the quark direction in the pair center-of-mass frame,
$\hat n$ is normal to the production plane, and $\hat r=\hat n\times\hat k$. The correlation matrix $C$ can be extracted from the cross section before the final-state spin indices are summed. For the class of density matrices considered below, the
Peres--Horodecki positive-partial-transpose (PPT)
criterion~\cite{Peres:1996dw,Horodecki:1997vt} can be conveniently
expressed in terms of the two quantities
\begin{align}
 \Delta_1 &=
 \sqrt{(C_{rr}-C_{kk})^2+(C_{rk}+C_{kr})^2}-1+C_{nn},
 \nonumber\\
 \Delta_2 &=
 \sqrt{(C_{rr}+C_{kk})^2+(C_{rk}-C_{kr})^2}-1-C_{nn}.
 \label{eq:deltas}
\end{align}
The state is entangled if and only if at least one of these quantities is
strictly positive. The amount of entanglement can then be quantified by the
concurrence, $\mathcal C[\rho]
 =\max\left\{ \Delta_1 /2 , \Delta_2/2, 0 \right\}$.
Bell nonlocality is tested through the CHSH inequality
\cite{Bell:1964kc,Clauser:1969ny}.  If $\mu_1\geq\mu_2\geq\mu_3$ are the
eigenvalues of $C^{\rm T}C$, the state violates CHSH precisely when
\cite{Horodecki:1995nsk}
\begin{equation}
 \mu_1+\mu_2>1,                                                \label{eq:chsh}
\end{equation}
with the algebraic range ending at $\mu_1+\mu_2=2$.  Entanglement and Bell
nonlocality are therefore related, but they are not equivalent for a generic
mixed state.

\section{Spin-density matrix in coherent diffraction}

Coherent diffraction provides a particularly clean laboratory for studying
the QCD Pomeron and the onset of gluon saturation: the target remains intact,
while the interaction is mediated by a color-singlet exchange. This has
motivated the development of increasingly observable quantities. Examples
range from the spin-density-matrix elements in exclusive
vector-meson production~\cite{Boussarie:2024pax,Boussarie:2026qzy}, to diffractive dijet
correlations proposed as probes of small-$x$ gluon tomography~\cite{Hatta:2016dxp,Mantysaari:2020lhf}. These studies show that information beyond the unpolarized cross section---such as azimuthal correlations, and helicity transitions---can reveal
otherwise hidden properties of the color-singlet exchange. Spin correlations between distinct particles in the final state offer a
further possibility, giving access to genuine bipartite quantum correlations. 

In Ref.~\cite{Fucilla:2025kit}, we constructed this matrix for a heavy
quark-antiquark pair produced in coherent diffractive DIS or UPC. This construction makes it possible to
investigate how the QCD Pomeron imprints itself on the entanglement structure of the final state.  In diffractive DIS, a virtual photon splits into a dipole with
longitudinal fractions $z$ and $\bar z=1-z$; the dipole then scatters
elastically through the color-singlet target amplitude $T(\bm p)$.  Two scalar
convolutions contain the target dependence,
\begin{equation}
 T_1=\int\!\frac{d^2\bm p\,T(\bm p)}{(\kk-\bm p)^2+\mu^2},\qquad
 T_2=-\frac{1}{k_\perp^2}\int\!\frac{d^2\bm p\,
 \kk\!\cdot\!\bm p\,T(\bm p)}{(\kk-\bm p)^2+\mu^2},
 \quad \mu^2=z\bar z Q^2+m^2.                               \label{eq:T12}
\end{equation}
The spin basis is most transparent in the pair center-of-mass frame, where we consider the set of variables
\begin{equation}
 \beta=\sqrt{1-\frac{4m^2}{M^2}},\qquad
 |\vec k|=\frac{M\beta}{2},\qquad
 \cos\theta=\frac{(2z-1)M}{\sqrt{M^2-4m^2}}.                 \label{eq:kin}
\end{equation}

For a longitudinal photon, all dependence on $T_1$ cancels after normalizing
the density matrix.  The remaining correlation matrix is orthogonal,
$(C^L)^{\rm T}C^L=\mathbb I$.  Hence $\mu_1+\mu_2=2$ for every allowed
$\beta$ and $\theta$: the pair is pure, maximally entangled, and saturates the
quantum CHSH bound.  This agrees with the one-gluon result
\cite{Qi:2025onf}, although in diffraction the $t$-channel state is a
color-singlet superposition of multiple gluon exchanges.

The transverse-photon density matrix is mixed in general and retains the
target-dependent ratio of $T_1$ and $T_2$.  Nevertheless, its coefficients
obey two identities,
\begin{align}
 &(C^T_{rr})^2+(C^T_{rk})^2+(C^T_{kr})^2+(C^T_{kk})^2
 -(C^T_{nn})^2=1,\nonumber\\
 &C^T_{nn}=-C^T_{rr}C^T_{kk}+C^T_{rk}C^T_{kr}.                \label{eq:identities}
\end{align}
Inserted into Eq.~\eqref{eq:deltas}, they imply the compact result
\begin{equation}
 \Delta_2=-\Delta_1=-2C^T_{nn},\qquad
 C^T_{nn}=-\frac{k_\perp^2(T_1+T_2)^2}
 {k_\perp^2(T_1+T_2)^2+2m^2T_1^2}\leq0.                    \label{eq:diff-result}
\end{equation}
Except at kinematic boundaries, the transverse state is thus entangled.  The
same calculation gives the spectrum of $(C^T)^{\rm T}C^T$ as
$\{1,(C^T_{nn})^2,(C^T_{nn})^2\}$, so that any nonzero concurrence is
accompanied by CHSH violation.  This equivalence is stronger than Gisin's
theorem for pure two-qubit states~\cite{Gisin:1991vpb}, because it holds here for a continuous family of mixed states.

\begin{figure}[t]
\centering
\begin{tikzpicture}[x=1cm,y=1cm,font=\small]
  \begin{scope}[shift={(0,0)}]
  \draw (0,0) ellipse (4 and 2.70);
    \draw (0,0) ellipse (3.4 and 2.10);
    \draw (0,0) ellipse (2.8 and 1.5);
    \draw (0,0) ellipse (2.2 and 0.9);
    \draw[fill=gray!20] (0,-0.10) ellipse (1.60 and 0.42);
    \node at (0,3.22) {Generic two-qubit states};
    \node at (0,2.32) {separable};
    \node at (0,1.72) {entangled};
    \node at (0,1.15) {Bell nonlocal};
    \node at (0,0.55) {pure \& entangled};
    \node at (0,-0.10) {maximally entangled};
  \end{scope}
  \begin{scope}[shift={(8.0,0)}]
    \draw (0,0) ellipse (2.9 and 1.75);
    \draw[fill=gray!20] (0,-0.20) ellipse (2.3 and 0.85);
    \node[text width=3.8cm,align=center] at (0,1.05)
      {entangled \& Bell nonlocal};
    \node[text width=5 cm,align=center] at (0,-0.20)
      {pure \& maximally entangled};
    \node at (-0.2,3.22) {Diffractive $\gamma_T^*\!\to\qqbar$};
  \end{scope}
\end{tikzpicture}
\caption{Comparison of the hierarchy of correlations in a
generic two-qubit system (left) and in the diffractive pair production
(right).  For a longitudinal photon, only the pure maximally entangled class exists.}
\label{fig:venn}
\end{figure}
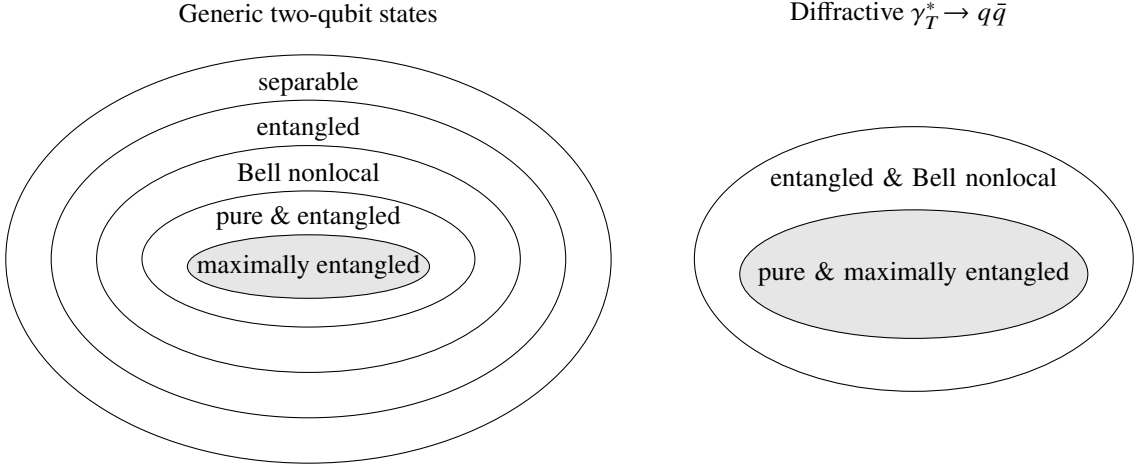

There is also a maximally entangled point.  Color transparency implies
$\int d^2\bm p\,T(\bm p)=0$.  Expanding Eq.~\eqref{eq:T12} around $z=1/2$,
\begin{equation}
 T_1\simeq\frac{k_\perp^2-\mu^2}{(k_\perp^2+\mu^2)^3}
 \int d^2\bm p\,p_\perp^2T(\bm p).                          \label{eq:zero}
\end{equation}
Thus $T_1$ crosses zero at $k_\perp\simeq\mu$; in photoproduction this is
$k_\perp\simeq m$.  At that point $C^T_{nn}=-1$, the state becomes pure and
maximally entangled, and CHSH violation is maximal.  The logical organization
of states is summarized in Fig.~\ref{fig:venn}.  A treatment based on gluon
generalized parton distributions leads instead to a richer target-dependent
pattern~\cite{Hatta:2025obw}, making the simple structure in
Eq.~\eqref{eq:diff-result} a characteristic result of the high-energy
color-singlet limit.

\section{Inclusive back-to-back production at small $x$}

Having discussed coherent diffraction, we now turn to inclusive
quark-antiquark production, $\gamma^* A\to q\bar q X$, in which one sums over
the possible final states of the target. We
focus on the back-to-back, or correlation, limit
$P_\perp\sim|\bm p_q|\sim|\bm p_{\bar q}|\gg
q_\perp=|\bm p_q+\bm p_{\bar q}|$, where the Color Glass Condensate expression for
DIS dijets~\cite{Gelis:2010nm,Dominguez:2011wm} reduces to a hard tensor
contracted with the Weizs\"acker--Williams (WW) gluon distribution.  Keeping
the spin indices open gives schematically
\begin{align}
 \frac{d\sigma^{L/T}_{\alpha\alpha'\beta\beta'}}
 {dz\,d^2\PP\,d^2\qq}
 &\propto
 \left.\frac{\partial\psi^{L/T*}_{\beta\beta'}}
 {\partial P^{\prime i}}
 \frac{\partial\psi^{L/T}_{\alpha\alpha'}}{\partial P^j}
 \right|_{\PP'=\PP} G^{ij}(\qq),\nonumber\\
 G^{ij}(\qq)&=\frac{\delta^{ij}}2G_0(q_\perp)
 +\left(\frac{q^iq^j}{q_\perp^2}-\frac{\delta^{ij}}2\right)G_2(q_\perp).
                                                               \label{eq:ktfact}
\end{align}
$G_0$ is the unpolarized WW distribution, while $G_2$ describes linearly
polarized gluons.  The hard tensor has the same scalar
and Pauli-matrix decomposition as Eq.~\eqref{eq:rho}.  Combining transverse
and longitudinal photons with polarization parameter $\varepsilon$ yields
\begin{align}
 \mathcal A={}&G_0(A_T^{(0)}+\varepsilon A_L^{(0)})
 +\cos(2\phi_{P,q})G_2(A_T^{(2)}+\varepsilon A_L^{(2)}),\nonumber\\
 \mathcal C_{ab}={}&\frac{1}{\mathcal A}\bigl[
 G_0(\widetilde C^{(0)}_T+\varepsilon\widetilde C^{(0)}_L)_{ab}
 +\cos(2\phi_{P,q})G_2
 (\widetilde C^{(2)}_T+\varepsilon\widetilde C^{(2)}_L)_{ab}\bigr].
                                                               \label{eq:inclusiveC}
\end{align}
This formula isolates a central difference from the diffractive channel: the normalized density matrix itself becomes a probe of the target only for a $\phi_{P,q}$-differential observable.

If one integrates over $\phi_{P,q}$, the $G_2$ term disappears and $G_0$
cancels in the normalized matrix.  The result reduces to the collinear
one-gluon density matrix studied by one of us in Ref.~\cite{Qi:2025onf}.  The transverse and
longitudinal components then compete, producing separable, entangled, and
Bell-nonlocal domains as the pair velocity $\beta$, the scattering angle, and
$\alpha=Q^2/M^2$ are varied.  This is already less rigid than coherent
diffraction.

\begin{figure}[t]
 \centering
 \includegraphics[width=0.96\textwidth]{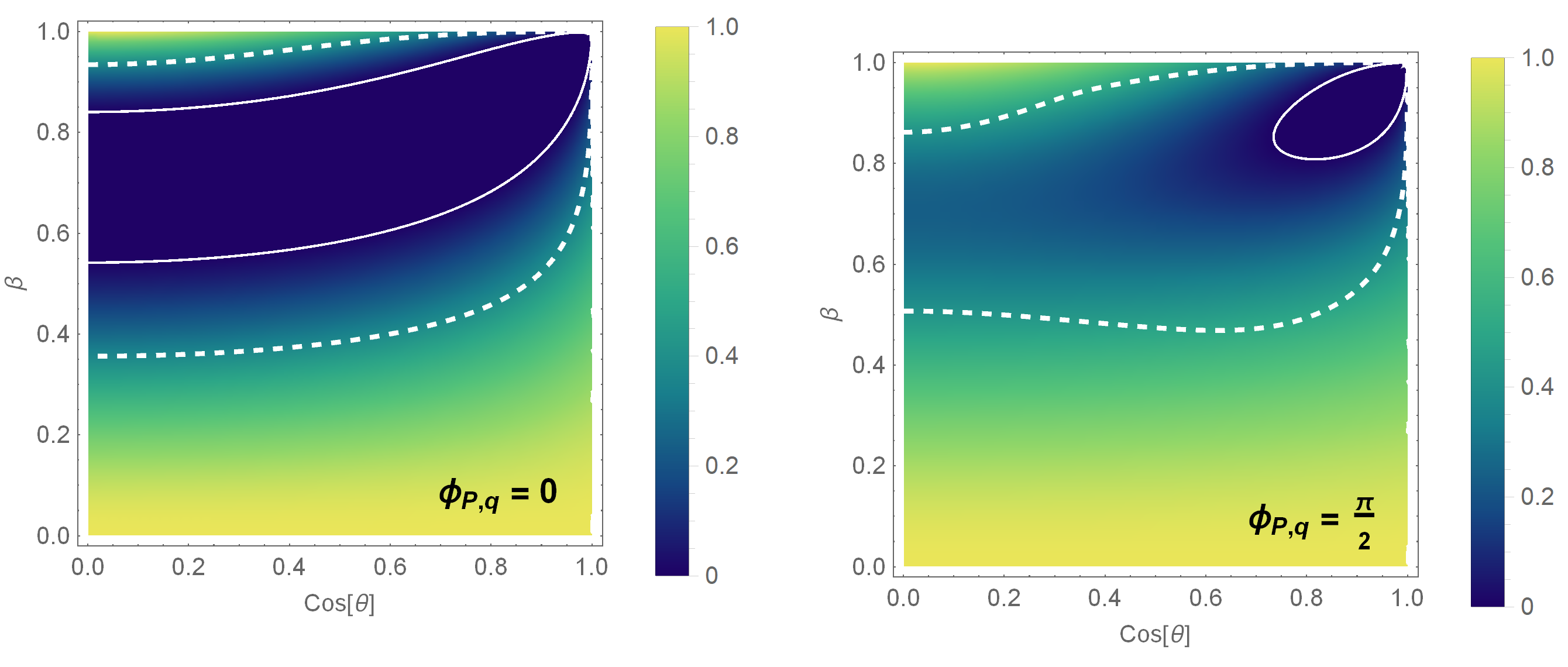}
 \caption{Concurrence of a heavy pair for $\phi_{P,q}=0$ (left) and
 $\phi_{P,q}=\pi/2$ (right), as a function of $\cos\theta$ and $\beta$, for a
 representative saturated target at $\alpha=Q^2/M^2=0$ and
 $q_\perp/Q_s=4$.  The solid curves enclose the separable domain, while the
 dashed curves delimit the larger domain without Bell nonlocality.  The color
 scale runs from zero (dark) to one (yellow).  Reduced two-panel selection
 adapted from Ref.~\cite{Fucilla:2026mkg}.}
 \label{fig:phi}
\end{figure}

The full azimuthal dependence reveals new information.  In a large nucleus or
at sufficiently small $x$, the McLerran--Venugopalan model
\cite{McLerran:1993ni} gives both $G_0$ and $G_2$ from the same saturated
classical field; the ratio $G_2/G_0$ grows away from $q_\perp=0$
\cite{Dumitru:2016jku}.  Equation~\eqref{eq:inclusiveC} then predicts only a
small change when $\cos(2\phi_{P,q})>0$, but a visible enhancement of the
concurrence for $\pi/4<\phi_{P,q}<3\pi/4$.  The effect is maximal near
$\phi_{P,q}=\pi/2$, where the relative momentum and dijet imbalance are
orthogonal.  Figure~\ref{fig:phi} displays a reduced comparison of the parallel
and orthogonal configurations at $q_{\perp}/ Q_s = 4$.  The solid contours delimit separability,
whereas the dashed contours delimit Bell-nonlocality.  The shrinking of the
dark separable region at $\phi_{P,q}=\pi/2$ is the clearest signature of the
linearly polarized WW distribution in this quantum-information observable. \\

In the dilute (BFKL) regime, the Weizs\"acker--Williams gluon correlator is maximally linearly polarized, implying $G_2(q_\perp)=G_0(q_\perp)$. At fixed hard kinematics, the resulting modulation of the concurrence is therefore maximal and approximately independent of the magnitude of the transverse-momentum imbalance. Multiple scattering in the saturation regime breaks this equality and suppresses $G_2/G_0$ around and below the saturation scale. Consequently, a suppression of the concurrence modulation at
$q_\perp\sim Q_s$, followed by its recovery in the dilute tail $q_\perp\gg Q_s$, could provide a novel signature of gluon saturation.

\section{From quark spins to measurable correlations}

The density matrix derived here describes the spins of the produced quark and antiquark at the partonic level. Experimentally, however, these spins must be reconstructed indirectly from the hadrons into which the heavy quarks fragment. A promising strategy is to select heavy-flavoured baryons, which can retain information about the parent-quark polarization, and use the angular distributions of their semileptonic weak decays as spin analysers. Heavy baryons can
retain a calculable fraction of the parent-quark polarization
\cite{Galanti:2015pqa}, and dedicated studies have assessed the prospects in
$b\bar b$ and $c\bar c$ samples~\cite{Kats:2023zxb}.  If $r_T$ denotes the
transverse polarization-retention factor and $\alpha_\pm$ the analyzing powers
of charge-conjugate decay products, the normal-normal correlation appears as
\begin{equation}
 \frac1\sigma\frac{d\sigma}{d\cos\theta_+d\cos\theta_-}
 =\frac14\left[1+\alpha_+\alpha_-r_T^2
 C_{nn}\cos\theta_+\cos\theta_-\right].                     \label{eq:decay}
\end{equation}
This makes the connection to data explicit, but the fragmentation fraction,
decay branching ratios, reconstruction efficiency, and spin transfer all
reduce the effective statistics.  Other approaches (see e.g.~\cite{Cheng:2025cuv}) may eventually provide complementary access without relying exclusively on
weak heavy-baryon decays. For experimental
analysis at EIC and LHC, the immediate priority is therefore to propagate the partonic
density matrix through fragmentation and detector-level reconstruction, to make a full feasibility study.

\section{Conclusions}

The spin-entanglement structure of final state particles offers a new perspective on the dynamics of high-energy QCD. In coherent diffraction, color-singlet exchange produces a
strikingly constrained state: longitudinal photons give maximal entanglement,
while transverse photons generate entanglement and Bell nonlocality together,
including a model-independent maximally entangled point tied to color
transparency.  Inclusive back-to-back production is less universal but more
phenomenologically abundant.  Its density matrix couples to $G_0$ and to the
linearly polarized WW distribution $G_2$, whose characteristic
$\cos2\phi_{P,q}$ dependence can enhance entanglement in orthogonal dijet
configurations. The next step is a realistic bridge from
quark-level correlations to reconstructed heavy-hadron observables, including
spin transfer, radiative corrections and experimental
acceptance.  Such studies will determine whether the clean theoretical
patterns described here can become practical measurements at the EIC or in UPCs at the LHC.

\end{document}